\documentclass{aa}

\usepackage{graphicx}
\usepackage{txfonts}
\usepackage{natbib}

\def\new{\relax}
\def\newer{\relax}

\begin{document}

\title{Orbits and masses in the young triple system TWA 5%
\thanks{Based on observations made with ESO Telescopes at the La Silla Paranal
  Observatory under programme ID 079.C-0103, 081.C-0393, 386.C-0205,
  087.C-0209, 088.C-0046, 089.C-0167, and 090.C-0184.}}

\author{R. K\"ohler\inst{1}
	\and
	T. Ratzka\inst{2}
	\and
	M.~G. Petr-Gotzens\inst{3}
        \and
        S. Correia\inst{4}
}

\institute{%
	Max-Planck-Institut f\"ur Astronomie, K\"onigstuhl 17,
	69117 Heidelberg, Germany,
        \email{koehler@mpia.de}
\and
	Universit\"ats-Sternwarte M\"unchen,
	Ludwig-Maximilians-Universit\"at,
	Scheinerstr. 1, 81679 M\"unchen, Germany
\and
	European Southern Observatory,
	Karl-Schwarzschild-Str.\ 2, 85748 Garching bei M\"unchen, Germany
\and
	Institute for Astronomy, University of Hawaii, 34 Ohia Ku Str.,
        Pukalani, HI 96768, USA
}

\date{Received 15 October 2012; accepted 25 August 2013}

\abstract{}{We aim to improve the orbital elements and determine the
  individual masses of the components in the triple system TWA\,5.}
{Five new relative astrometric positions in the H band were recorded
  with the adaptive optics system at the Very Large Telescope (VLT).
  We combine them with data from the literature and a measurement
  in the Ks band.  We derive an improved fit for the orbit of
  TWA\,5Aa-b around each other.
  Furthermore, we use the third component, TWA\,5B, as an astrometric
  reference to determine the motion of Aa and Ab around their center
  of mass and compute their mass ratio.
}
{We find an orbital period of $6.03\pm0.01\rm\,years$ and a semi-major
  axis of $63.7\pm0.2\rm\,mas$ ($3.2\pm0.1\rm\,AU$).
  With the trigonometric distance of $50.1\pm1.8\rm\,pc$, this yields a
  system mass of $0.9\pm0.1\,M_\odot$, where the error is dominated by
  the error of the distance.
  The dynamical mass agrees with the system mass predicted by a number
  of theoretical models if we assume that TWA5 is at the young end of
  the age range of the TW Hydrae association.
  We find a mass ratio of $M_{\rm Ab} / M_{\rm Aa} = 1.3^{+0.6}_{-0.4}$,
  where the less luminous component Ab is more massive.
  This result is likely to be a consequence of the large uncertainties
  due to the limited orbital coverage of the observations.
}{}

\keywords{Stars: low-mass --
	  brown dwarfs --
          Stars: fundamental parameters --
	  Stars: individual: TWA 5 --
	  Binaries: close --
	  Celestial Mechanics}

\maketitle


\section{Introduction}

The mass is the most important parameter for the structure and
evolution of a star. Therefore, empirical mass determinations are
crucial for our understanding of stellar astrophysics.
In particular, this is the case for low-mass pre-main-sequence (PMS)
stars and brown dwarfs, where a number of evolutionary models with
different mass predictions exist
\citep[e.g.,][]{DAntoMazzi1997,BCAH98,PallaStahler99,siess2000,tognelli2011}.
Binary stars are the {\it only\/} way to measure stellar masses
directly without relying on theoretical models.  They are therefore
valuable test cases for theoretical pre-main-sequence tracks.

The TW Hydrae association is one of the closest associations of young
stars with a median distance of 56\,pc \citep{weinberger2012}.
Its members have been shown to be young ($\sim$5 -- 15\,Myr),
based on lithium abundance tests and their positions in the H-R
diagram \citep[and references therein]{weintraub2000}.
The TW Hydrae association is therefore an ideal region for studying
spatially resolved PMS binaries.

The object \object{TWA\,5} is one of the five original members of the
TW~Hydrae association identified by \citet{Kastner1997}.  It is
composed of at least three components: a pair of low-mass stars;
\hbox{TWA\,5Aa-b}, which had a separation of 55\,mas when it was
discovered by \citet{Macintosh2001}; and a brown dwarf companion,
TWA\,5B, located about 2'' away \citep{webb1999}.  \citet{Konop2007}
presented an orbital solution for the inner binary with a period of
$5.94\pm0.09$\,years and a semi-major axis of $66\pm5$\,mas, which
results in a binary system mass of $0.71\pm0.14\,M_\odot$ (for a
distance of 44\,pc).

In this paper, we present new relative astrometric measurements
collected between 2007 and 2013.
The binary has completed two full orbits since its discovery in 2000
and one orbit since the work of \citet{Konop2007}.
We combine the new data with data from the literature to derive an
improved orbit solution.
{\new Furthermore, \citet{weinberger2012} published new results for the
  parallax of TWA\,5, which significantly changes and improves the
  mass estimates from orbit determinations.}


\begin{figure}[tp]
\centerline{%
  \includegraphics[width=0.5\hsize]{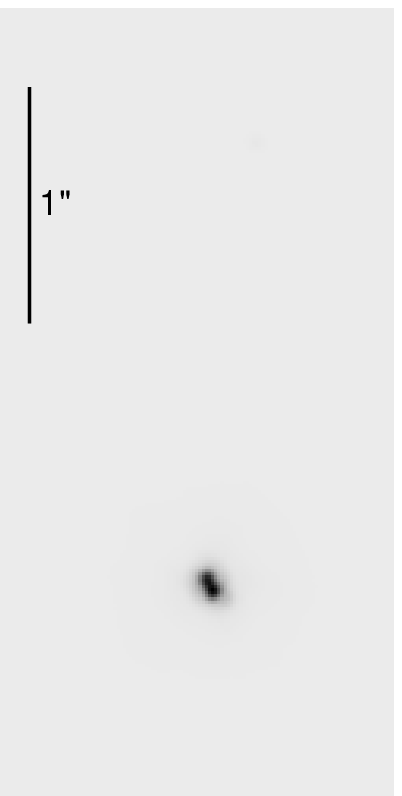}%
  \includegraphics[width=0.5\hsize]{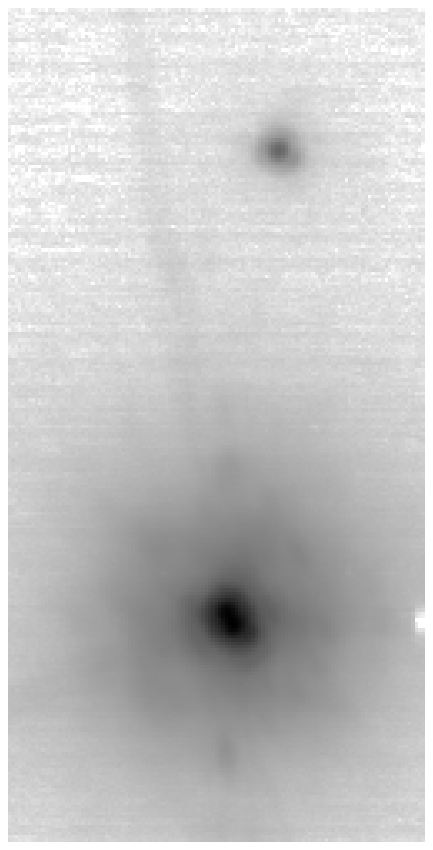}}
\caption{Image of TWA\,5Aa/b and TWA\,5B taken with NACO in January
  2012. North is up and east is to the left.
  Both panels show the same image with a linear scale on the left and
  with a logarithmic scale on the right to unveil the low-mass
  companion TWA\,5B.}
\label{NACOPic}
\end{figure}


\begin{figure}[tp]
\hbox to\hsize{%
  \includegraphics[height=0.5\hsize]{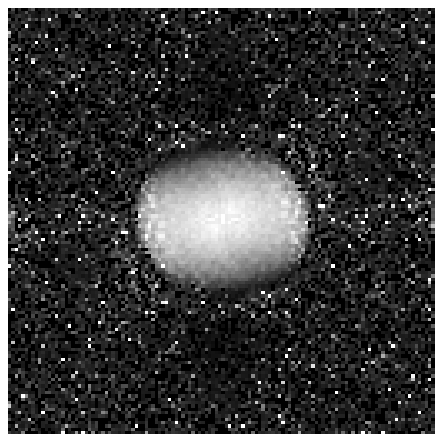}\hss%
  \includegraphics[height=0.5\hsize]{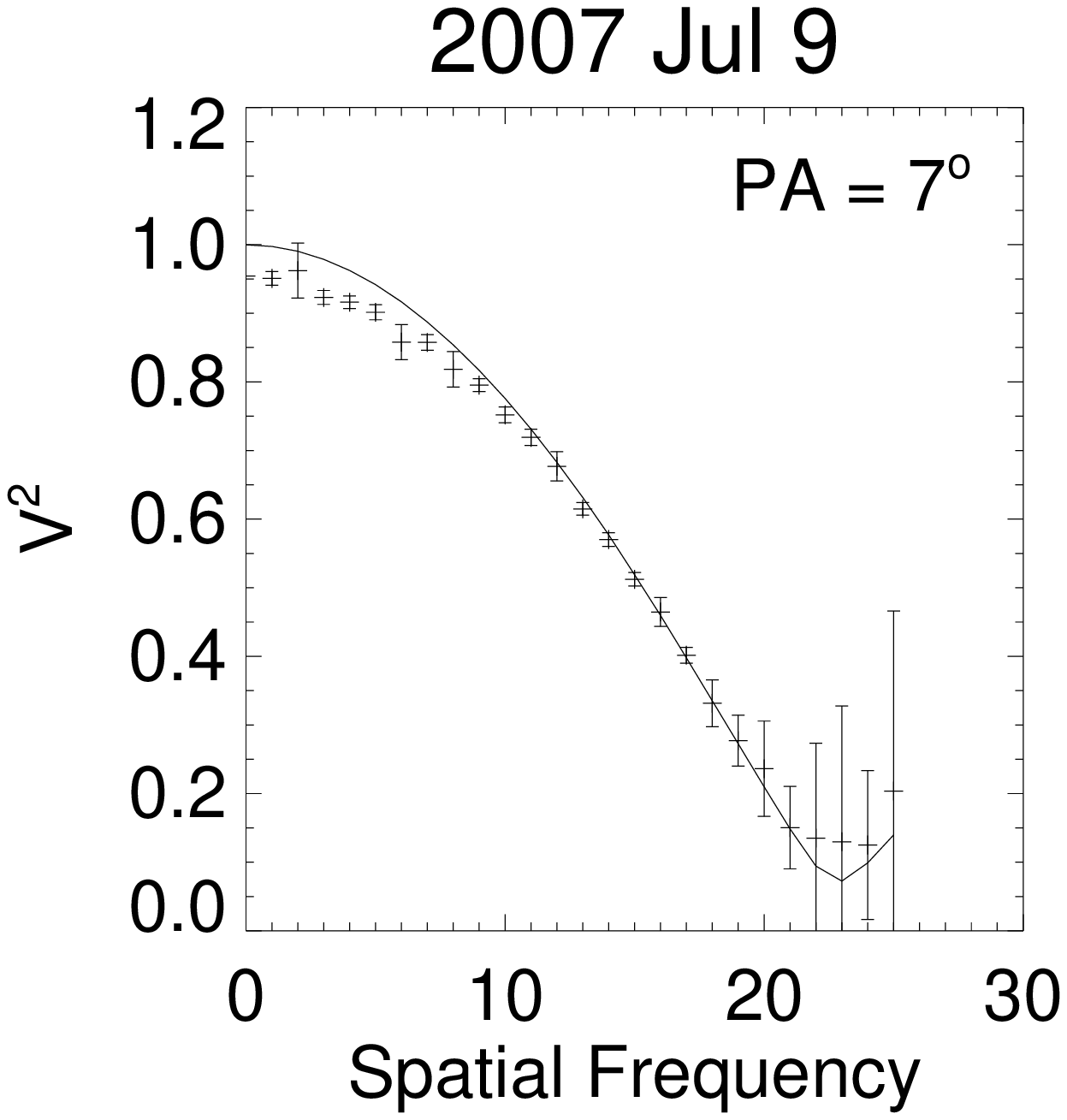}\hss%
}
\hbox to\hsize{%
  \includegraphics[height=0.5\hsize]{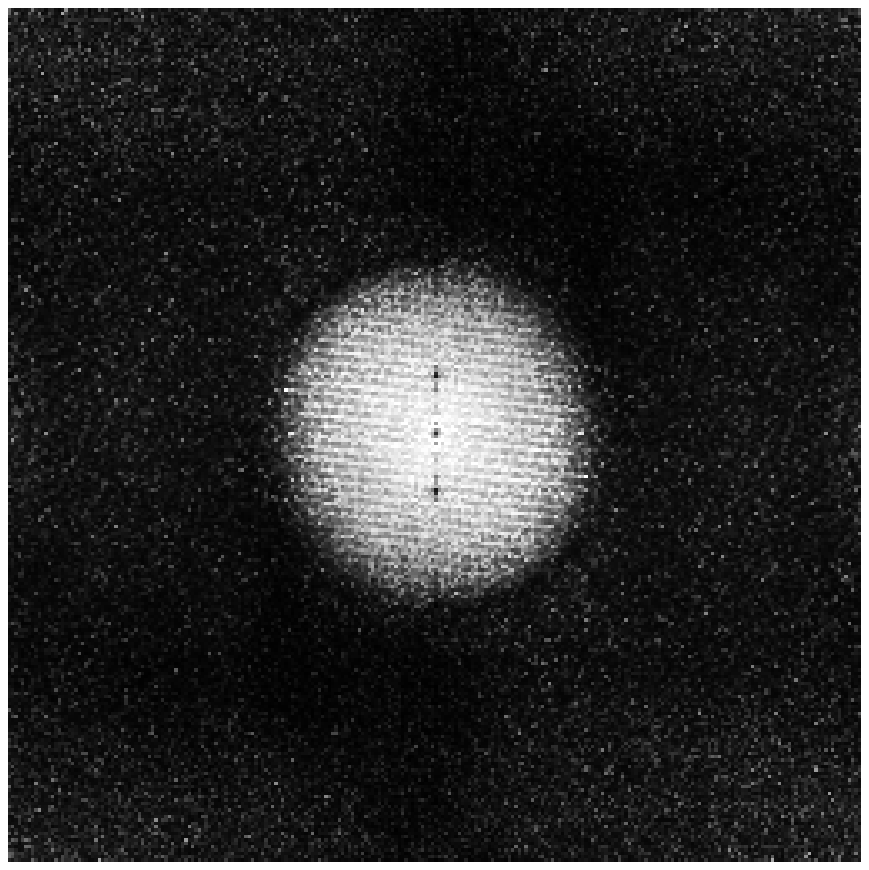}\hss%
  \includegraphics[height=0.5\hsize]{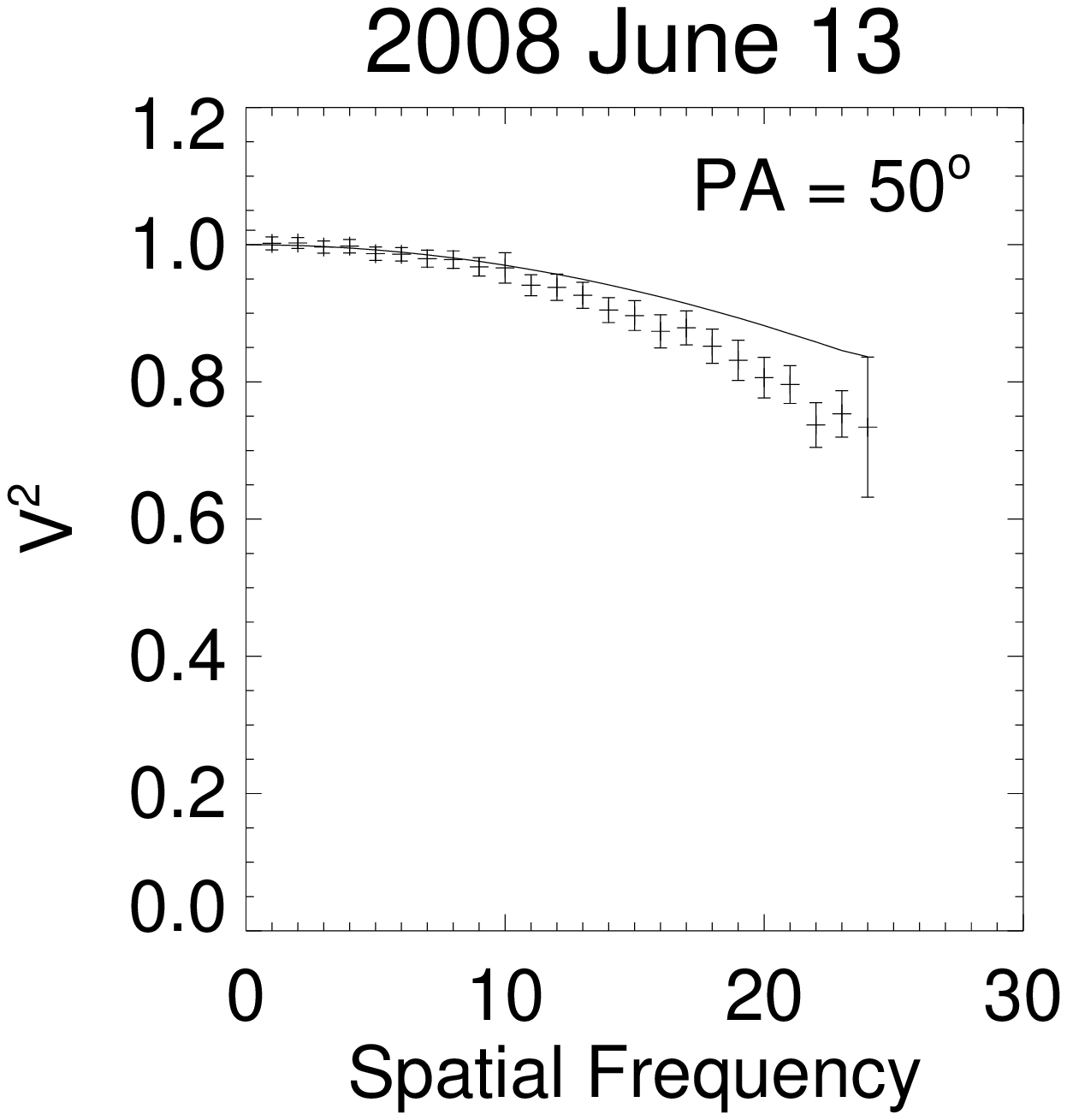}\hss%
}
\caption{Modulus of the visibility of TWA\,5Aa/b observed through the Ks
  filter in July 2007 and June 2008.  The data points show the
  measurements, and the line in the plot for 2007 is the result of a
  fit of a binary model to the data.
  The line in the plot for 2008 does not show a fit to the data but
  shows the visibility for the position computed from the orbit.
  The separation predicted at this date is too small to be resolved,
  which is confirmed by the data.}
\label{Vis2007to8pic}
\end{figure}


\begin{figure}[htp]
\hbox{%
 \includegraphics[height=0.5\hsize]{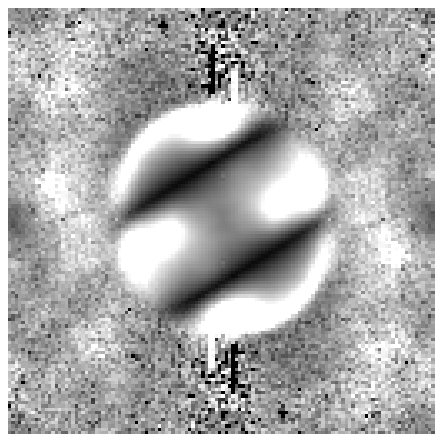}%
 \includegraphics[height=0.5\hsize]{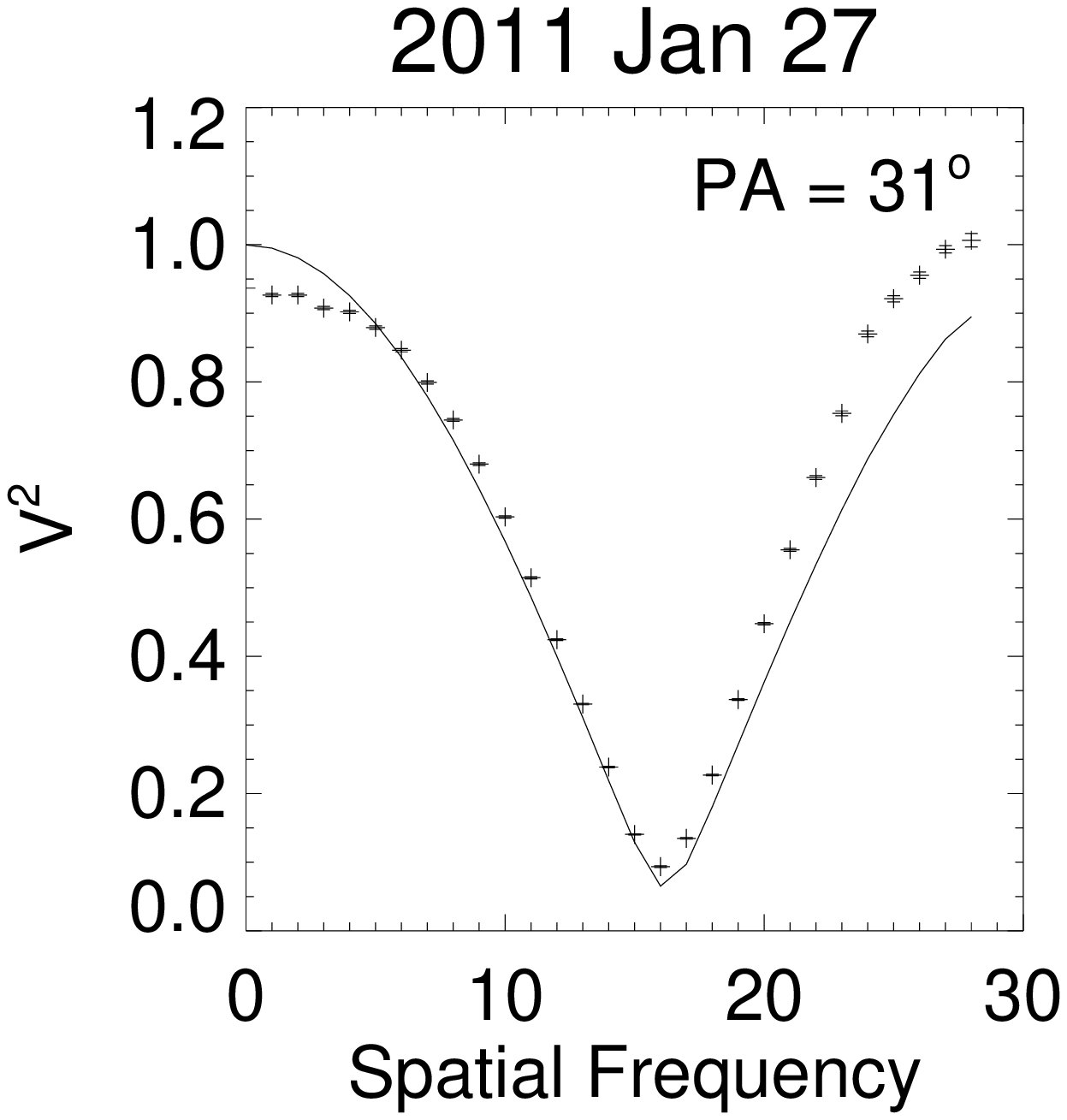}\hss}%
\hbox{%
 \includegraphics[height=0.5\hsize]{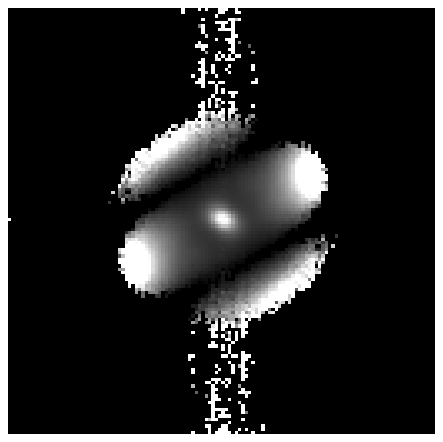}%
 \includegraphics[height=0.5\hsize]{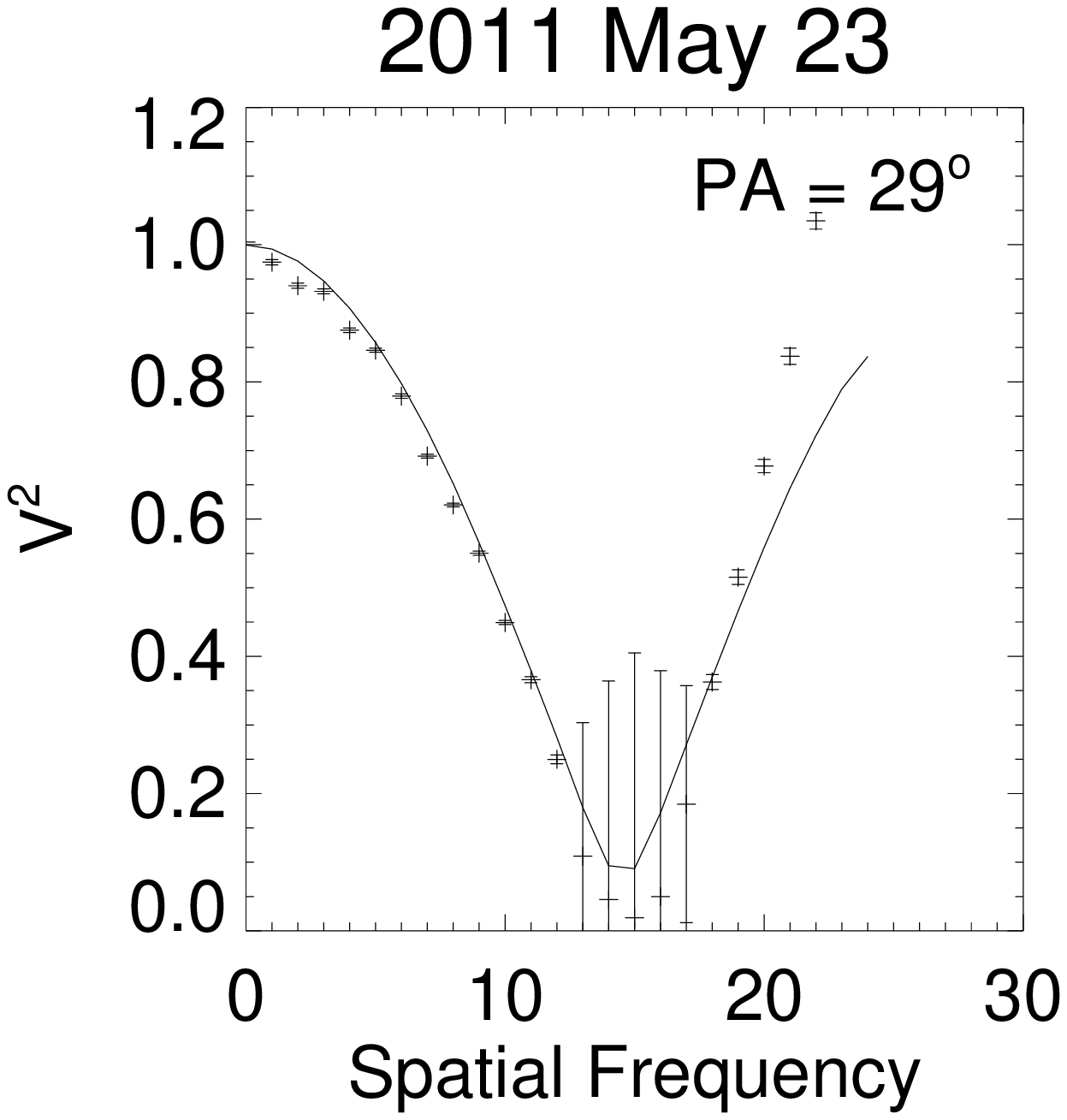}\hss}%
\hbox{%
 \includegraphics[height=0.5\hsize]{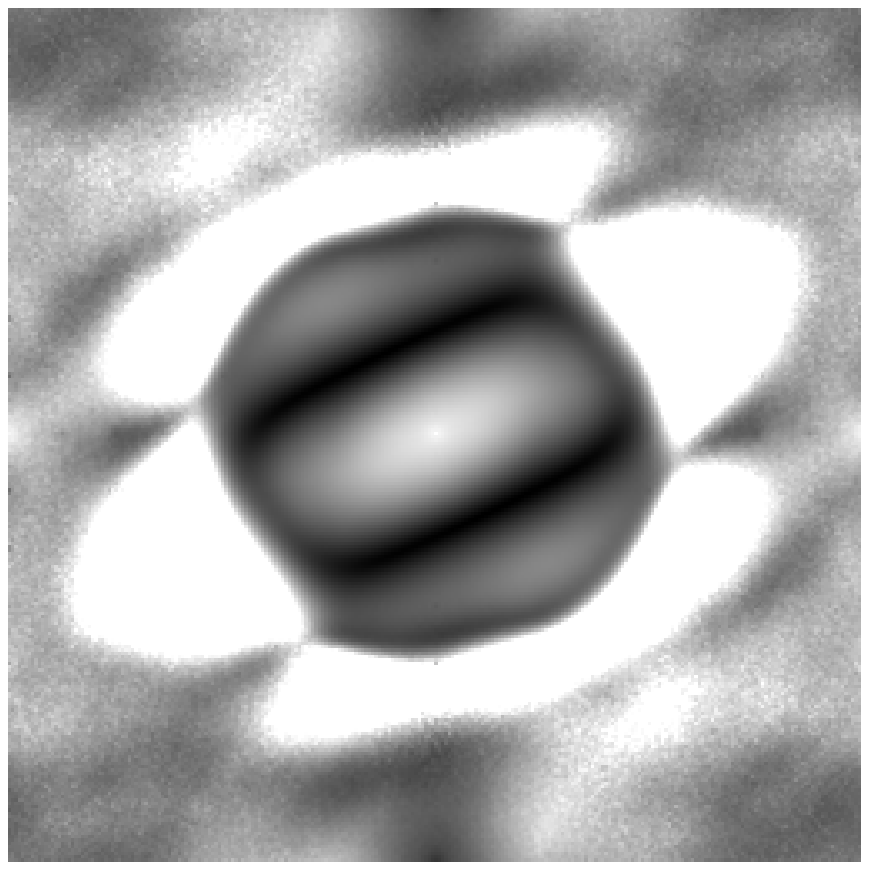}%
 \includegraphics[height=0.5\hsize]{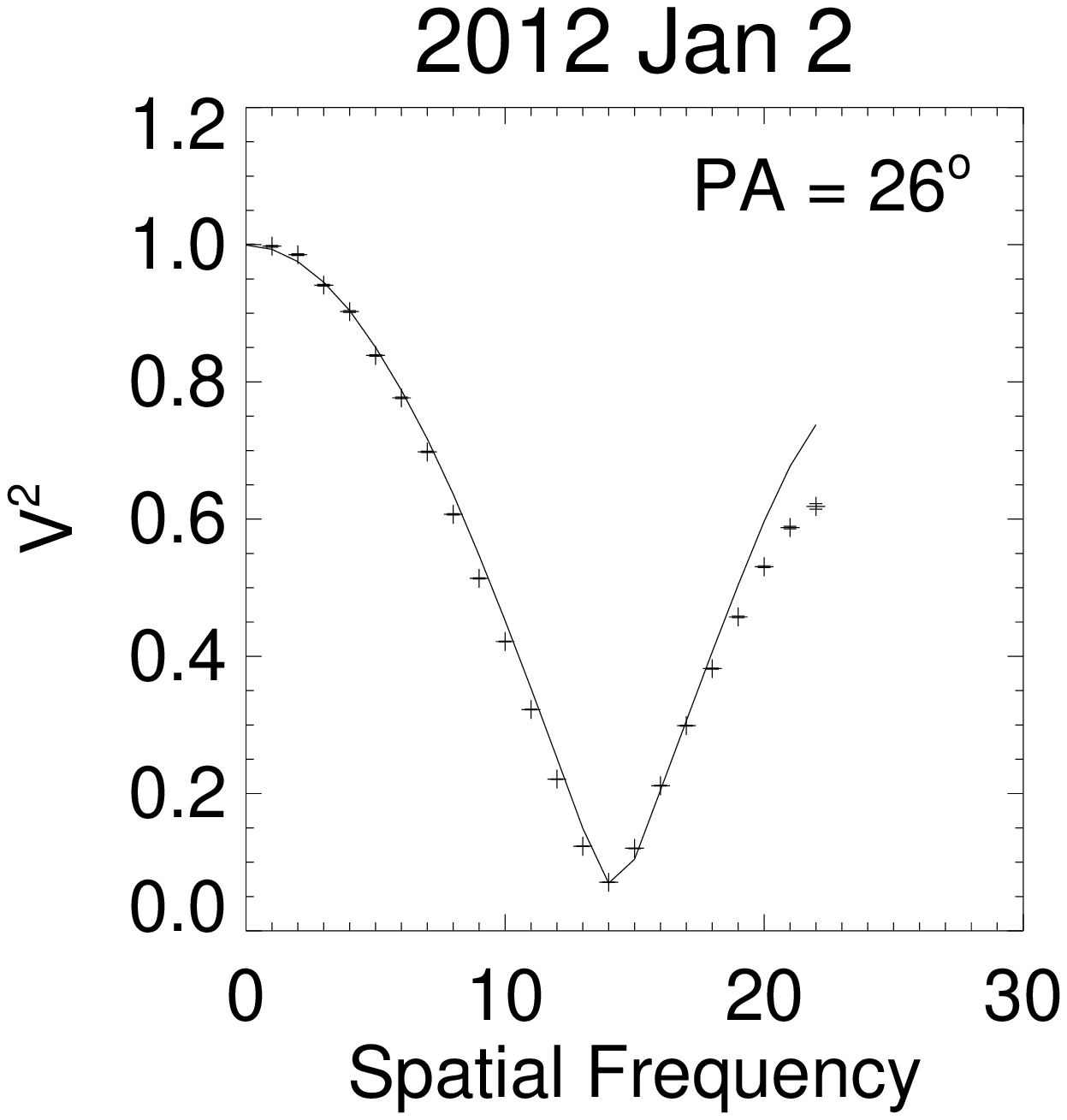}\hss}%
\hbox{%
 \includegraphics[height=0.5\hsize]{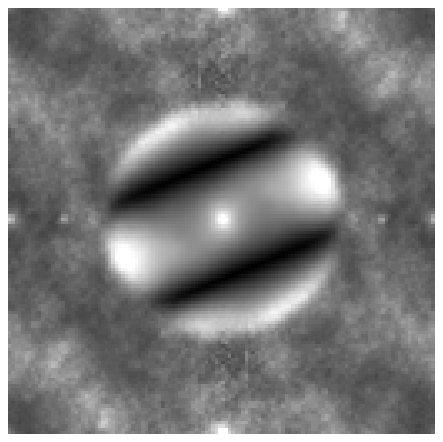}%
 \includegraphics[height=0.5\hsize]{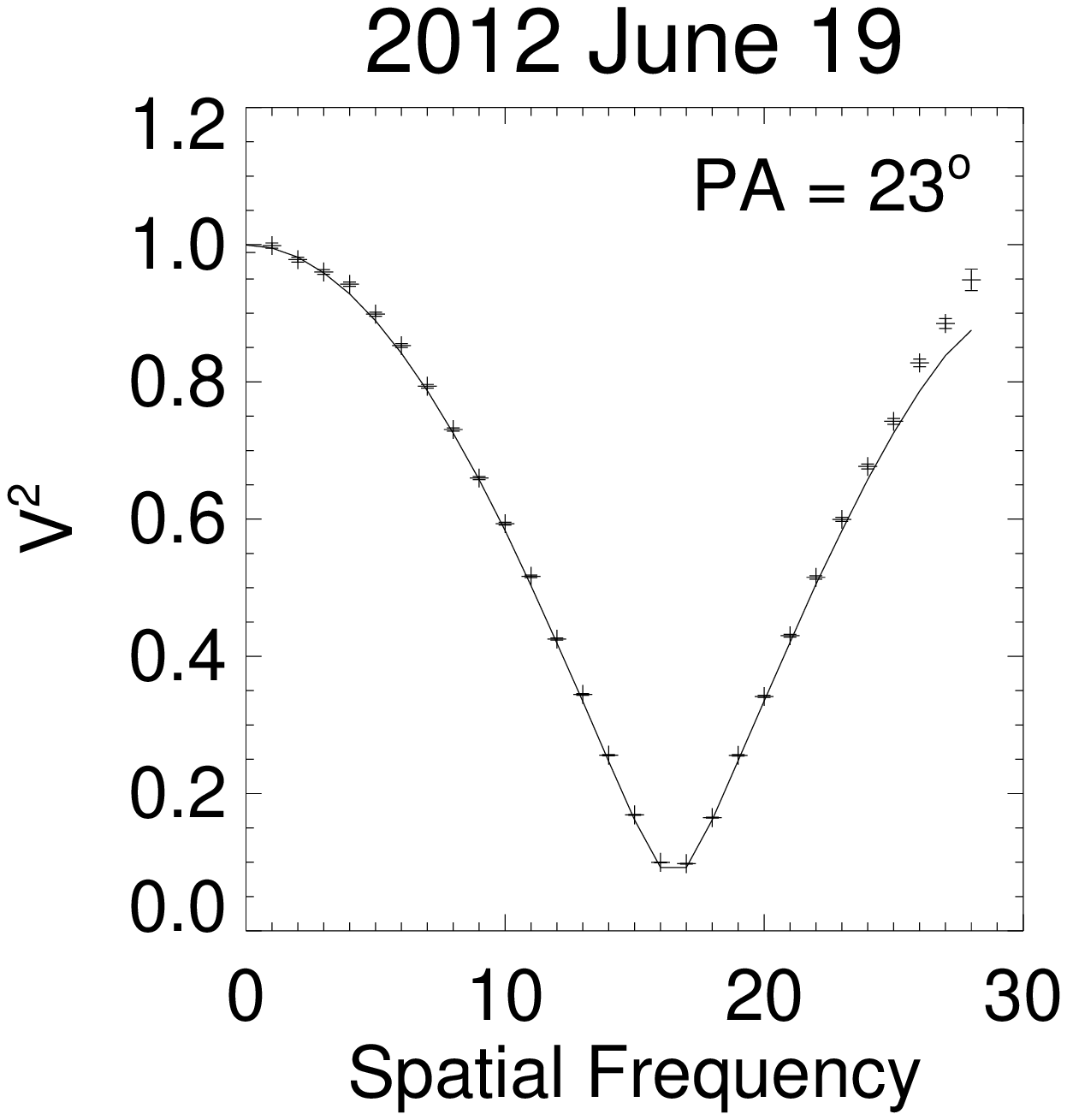}\hss}%
\hbox{%
 \includegraphics[height=0.5\hsize]{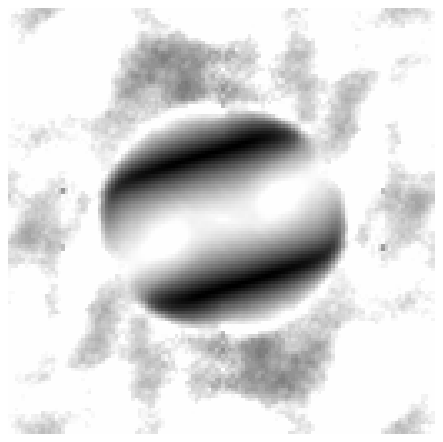}%
 \includegraphics[height=0.5\hsize]{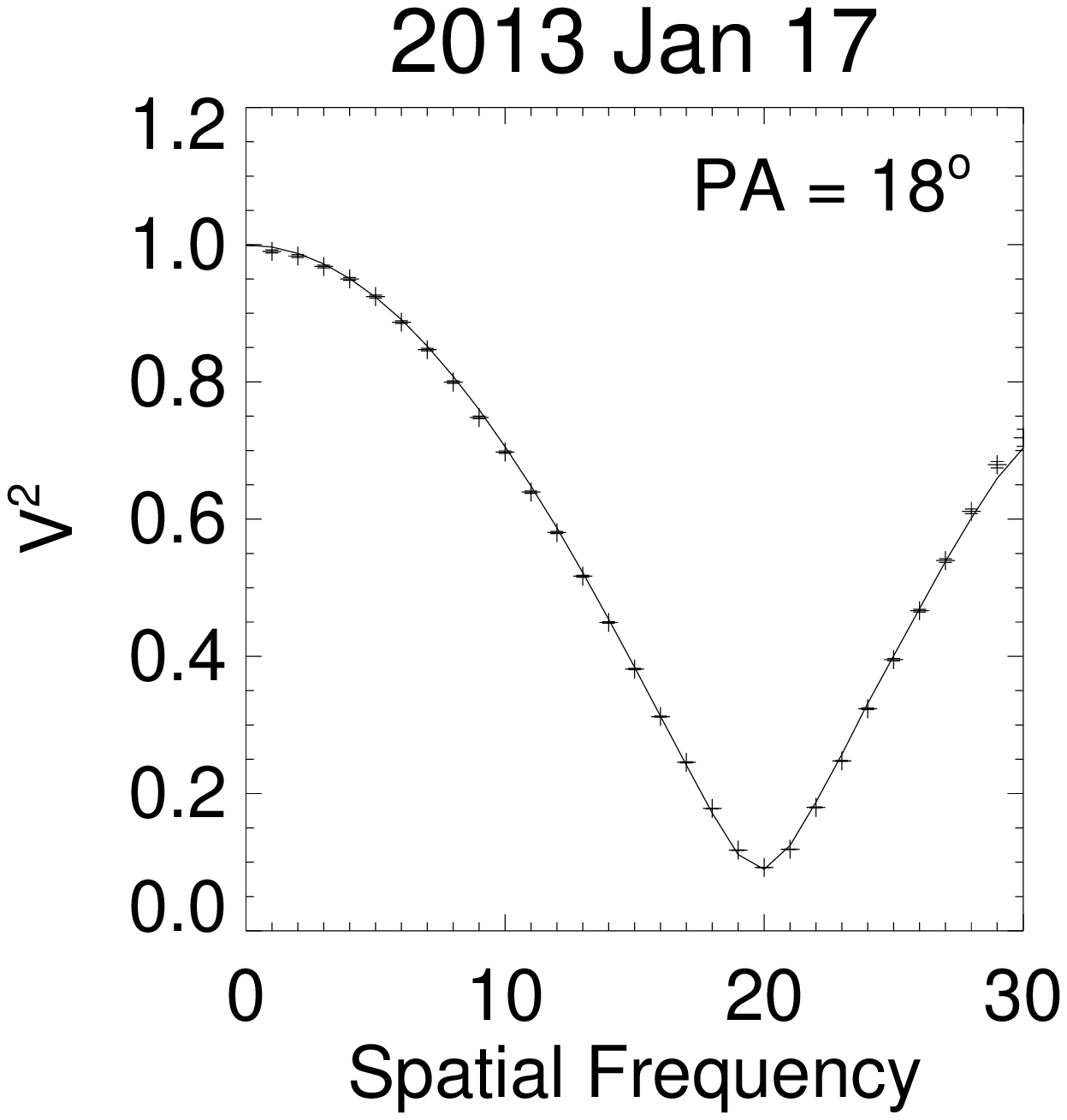}\hss}%
\caption{Modulus of the visibility of TWA\,5Aa/b observed through the H
  filter between 2011 and 2013. The data points show the
  measurements and the lines are the results of fits of binary models
  to the data.}
\label{Vis2011to13pic}
\end{figure}



\section{Observations and data reduction}

The first new observations were taken in 2007 and 2008 in the course
of a different project and have been retrieved from the archive of the
European Southern Observatory (ESO).
The data was recorded with NAOS/Conica (NACO for short), the adaptive
optics, near-infrared camera at the ESO Very Large Telescope (VLT) on
Cerro Paranal, Chile \citep{rousset03, lenzen03}.
The Ks photometric filter with a central wavelength of
$2.18\rm\,\mu m$ was used.  The resulting diffraction limit is
$\lambda/D = 56\rm\,mas$.

In 2011, we started monitoring the orbital motion of TWA\,5.
The observations were also carried out with NACO but in the H-band
filter with a central wavelength of $1.66\rm\,\mu m$, which gives a
smaller diffraction limit of $43\rm\,mas$.
We used the S13 camera, which provides a field of view of
$\sim13.6\arcsec\times13.6\arcsec$.
To allow the use of speckle-interferometric algorithms (see below),
we employed the cube-mode of CONICA and recorded many exposures with
short integration times of $0.2\rm\,s$ each.  One cube contained
typically $\sim$100 images. Eight cubes at different positions on the
detector chip were taken at each epoch.

The NACO images were sky subtracted with a median sky image, and
bad pixels were replaced by the median of the closest good neighbors.
Figure~\ref{NACOPic} shows an example of the result.
The separation of the binary is close to the diffraction limit, which
makes it difficult to disentangle the point spread functions of the
two components in the images.  Therefore, we used our software for
speckle interferometry \citep[see e.g.][]{Koehler2000}.
In this program, the modulus of the complex visibility (i.e.,\ the
Fourier transform of the object brightness distribution) is determined
from power spectrum analysis.  The phase is computed using the
Knox-Thompson algorithm \citep{KnoxThomp74} and from the bispectrum
\citep{Lohmann83}.
Examples of the results are presented in Figs.~\ref{Vis2007to8pic}
and~\ref{Vis2011to13pic}.

For the deconvolution of speckle images, an unresolved star is
required to calibrate the point spread function (PSF).
Initially, we used TWA\,5B for this purpose, since it is present in all
our frames and well separated from TWA\,5A so that the PSFs do not
overlap.
In principle, TWA\,5B is an ideal PSF reference, since it lies within
the isoplanatic patch and was observed simultaneously with TWA\,5A.
However, it is much fainter, which results in a high noise level in
the visibilities.

Starting with our own observations in 2011, we observed the binary
HIP\,56620 immediately after TWA\,5 to calibrate the pixel scale and
orientation.
We tried using the brighter component of HIP\,56620 to deconvolve the
speckle-images of TWA\,5A.  Although the observations were not intended
for this, we found that the reconstructed visibilities are
significantly less noisy than those based on TWA\,5B.
Therefore, we used the visibilities deconvolved with HIP\,56620 for
the astrometric measurement.

The final complex visibility was computed by averaging moduli and
phases derived from the eight data cubes.
The parameters of the binary (separation, position angle, and flux
ratio) are determined by a multidimensional least-squares fit of a
binary model to the complex visibility.
Uncertainties were estimated by fitting each of the individual
visibilities computed from one of the eight data cubes.
The standard deviation of the results were adopted as the error of the
relative positions.

In 2007, the separation of the binary was smaller than the diffraction
limit $\lambda/D = 56\rm\,mas$ of the observations but larger than
$\lambda/2D$.
This means that the first minimum of the visibility but not the
following maximum was measured (see Fig.~\ref{Vis2007to8pic}).
As a result, the position derived from these measurements has a larger
uncertainty than the other observations.

The visibilities measured in June 2008 show no clear sign of binarity.
This is not surprising, since the expected separation at the time of
the observations is about 14\,mas, which is significantly smaller than
$\lambda/2D$ in the Ks-band.
The line in the lower right panel in Fig.~\ref{Vis2007to8pic}
  shows the visibility of the binary at the position predicted by the
  orbit we derive in Sect.~\ref{InnerOrbitSect}.
  The data indicates that the binary is partially resolved and that
  the separation might be a bit larger than predicted.
  However, we do not regard this as a resolved measurement and do not
  use it in the orbit fit.

To measure the relative positions of TWA\,5A and B, we used the
{\tt starfinder} program \citep{Diolaiti00}.
The program does not recognize that TWA\,5A is a binary.
Since TWA\,5Aa and Ab have comparable brightness, the measured
position is not centered on the brighter component but refers to the
combined center of light of Aa and Ab.


\subsection{Astrometric calibration}
\label{AstCalSect}

{\new
Since the observations in 2007 and 2008 were not intended for
high-precision astrometry, no special measurements for the calibration
of the plate scale and orientation of the camera were taken.
We resorted to images of fields in the Orion Trapezium that had been
taken for a different project in October 2007.
These observations were done in the Ks filter, which is the same as
the observations of TWA\,5 in 2007 and 2008.
We reduced the images of the Trapezium in the same way as the images
of the science target.
The {\tt starfinder} program was used to measure the positions of the
cluster stars on the detector.
The pixel positions were compared to the coordinates given in
\cite{Close2012}, which in turn are based on HST-positions of a few
cluster stars.
We computed the mean pixel scale and orientation of NACO from a global
fit of all star positions.  The scatter of values derived from star
pairs were used to estimate the errors.
The pixel scale was determined as $(13.271\pm0.013)\rm\,mas/pixel$.
The position angle (PA) of the vertical axis of the detector was
$(0.43\pm0.10)^\circ$ (measured from north to east).

During our astrometric monitoring program in 2011 to 2013, images of
the Hipparcos-binary HIP\,56620 were taken.
However, we found that its relative position in January 2011 was
significantly different from the position measured by Hipparcos about
20 years earlier (see Appendix \ref{AstromApp}).
Furthermore, we found a systematic change of the position with time,
which led us to the conclusion that the relative position of
HIP\,56620 is not stable enough to serve for the astrometric
calibration of NACO.

Therefore, we decided to rely on the Trapezium cluster for the
calibration of all our observations.
Two data sets are available from the ESO archive that were recorded on
30 January 2011 and 2 January 2012.
The average pixel scale resulting from these data was
$(13.285\pm0.013)\rm\,mas/pixel$, and the average PA of the y-axis of
the images was $(0.81\pm0.1)^\circ$ (measured from north to
east)\footnote{\newer The difference in orientation measured in 2007
  can be explained by one or more technical interventions on NACO in
  2008 \citep{Kervella2013}.}.
These values were used to calibrate our measurements.
The errors of the calibration were added to the errors of the
measurements when we computed the relative positions.

The calibrated positions of TWA\,5Ab relative to TWA\,5Aa are listed in
Table~\ref{ObsTab} with data taken from the literature.
Table~\ref{ObsABTab} contains the positions of TWA\,5B relative to the
photocenter of A.
They agree well with the result of \citet{weinberger2012}.
}


\begin{table*}[ht]
\caption{Astrometric measurements of TWA\,5\,Aa-b.}
\label{ObsTab}
%
\begin{center}
\begin{tabular}{cccl@{${}\pm{}$}lr@{${}\pm{}$}lr@{${}\pm{}$}lccc}
\noalign{\vskip1pt\hrule\vskip1pt}
\multicolumn{2}{l}{Date and time (UT)} & Filter & \multicolumn{2}{c}{$d$ [mas]} & \multicolumn{2}{c}{PA~$[^\circ]$} & \multicolumn{2}{c}{Flux ratio} & Reference & $|\Delta d|/\sigma_d$ & $|\Delta\rm PA|/\sigma_{PA}$ \\
\noalign{\vskip1pt\hrule\vskip1pt}
2000 Feb 20 &  & H & $ 54.8$&$0.5$ & $ 25.9$&$1.0$ & $0.92$&$0.02$ & 1 &  1.4 &  0.7 \\
2000 Feb 22 &  & K' & $ 54.0$&$3.0$ & $ 24.2$&$3.0$ & $0.90$&$0.06$ & 2 &  0.5 &  0.3 \\
2001 May  6 &  & K & $ 35.1$&$0.2$ & $ 12.7$&$1.1$ & $0.79$&$0.06$ & 3 &  1.3 &  0.1 \\
2002 May 23 &  & K & $ 13.0$&$3.0$ & $313.7$&$3.0$ & $0.81$&$0.05$ & 3 &  0.3 &  3.5 \\
2003 Dec  5 &  & K & $ 30.6$&$0.4$ & $227.4$&$5.5$ & $0.82$&$0.03$ & 3 &  0.8 &  0.7 \\
2004 Dec 18 &  & K & $ 51.5$&$0.9$ & $ 32.1$&$2.2$ & $0.72$&$0.05$ & 3 &  1.2 &  0.4 \\
2005 Feb 16 &  & Fe {\sc ii} & $ 53.0$&$1.0$ & $ 32.6$&$5.2$ & $0.78$&$0.11$ & 3 &  1.4 &  0.2 \\
2005 May 27 &  & K & $ 57.4$&$0.3$ & $ 29.7$&$0.3$ & $0.81$&$0.03$ & 3 &  1.5 &  0.7 \\
2005 Dec 12 &  & H & $ 57.1$&$2.0$ & $ 28.9$&$1.0$ & $0.92$&$0.03$ & 3 &  0.2 &  2.3 \\
2007 Jul  9 & 23:22 & Ks & $ 36.6$&$4.0$ & $  7.5$&$2.0$ & $0.89$&$0.02$ & 4 &  1.3 &  1.2 \\
2011 Jan 27 & 04:36 & H & $ 52.9$&$0.3$ & $ 31.0$&$0.6$ & $0.90$&$0.02$ & 4 &  0.8 &  2.1 \\
 & 04:45 & J & \multicolumn{2}{c}{} & \multicolumn{2}{c}{} & $0.98$&$0.05$ & 4 \\
2011 May 23 & 00:26 & H & $ 59.9$&$2.4$ & $ 29.0$&$2.4$ & $0.92$&$0.05$ & 4 &  1.0 &  0.5 \\
 & 00:40 & J & \multicolumn{2}{c}{} & \multicolumn{2}{c}{} & $0.78$&$0.15$ & 4 \\
2012 Jan  2 & 07:29 & H & $ 60.6$&$0.5$ & $ 26.1$&$0.4$ & \multicolumn{2}{c}{---\tablefootmark{a}} & 4 &  6.4\tablefootmark{b} &  0.7\tablefootmark{b}\\
2012 Jun 19 & 22:54 & H & $ 51.7$&$0.3$ & $ 23.4$&$0.4$ & $0.86$&$0.01$ & 4 &  1.1 &  0.6 \\
 & 23:01 & J & \multicolumn{2}{c}{} & \multicolumn{2}{c}{} & $0.76$&$0.03$ & 4 \\
2013 Jan 17 & 07:05 & H & $ 42.4$&$0.3$ & $ 18.3$&$0.4$ & $0.83$&$0.01$ & 4 &  1.3 &  1.4 \\
 & 07:13 & J & \multicolumn{2}{c}{} & \multicolumn{2}{c}{} & $0.87$&$0.04$ & 4 \\
\noalign{\vskip1pt\hrule\vskip1pt}
\end{tabular}
\tablebib{(1)~\citet{Macintosh2001}; (2)~\citet{Brandeker2003}; (3)~\citet{Konop2007}; (4)~this work}
\tablefoottext{a}{Detector saturated}
\tablefoottext{b}{Not used for the orbit fit}
\end{center}
\end{table*}


\begin{table}[t]
\caption{Astrometric measurements of TWA\,5B relative to the
  photocenter of TWA\,5A.}
\label{ObsABTab}
%
\begin{center}
\begin{tabular}{lcl@{${}\pm{}$}lr@{${}\pm{}$}l}
Date (UT) &  Filter & \multicolumn{2}{c}{$d$ [mas]} & \multicolumn{2}{c}{PA~$[^\circ]$}\\
\noalign{\vskip1pt\hrule\vskip1pt}
2007 Jul  9 & K & $1901.6$&$2.0$ & $356.4$&$0.2$\\
2011 Jan 27 & H & $1887.5$&$7.4$ & $355.2$&$0.2$\\
2012 Jan  2 & H & $1878.8$&$2.3$ & $355.0$&$0.1$\\
2012 Jun 19 & H & $1874.6$&$2.5$ & $354.8$&$0.1$\\
2013 Jan 17 & H & $1872.7$&$2.0$ & $354.5$&$0.1$\\
\noalign{\vskip1pt\hrule\vskip1pt}
\end{tabular}
\end{center}
\end{table}


\section{The orbit of TWA\,5Aa-b}
\label{InnerOrbitSect}

We estimated the orbital parameters of TWA\,5Aa-b by fitting orbit
models to the observations listed in Table~\ref{ObsTab}.
We used a grid-search for eccentricity $e$, period $P$, and time of
periastron $T_0$, a procedure used previously
\citep[e.g.,][]{Koehler2008,Koehler2012}.
At each grid point, the Thiele-Innes elements were determined
by a linear fit to the observational data using singular value
decomposition.  From the Thiele-Innes elements, the semi-major axis
$a$, the angle between node and periastron $\omega$, the position
angle of the line of nodes $\Omega$, and the inclination $i$ were
computed.

Since the orbit of TWA\,5Aa-b is already well known \citep{Konop2007},
only a small range of parameter values had to be scanned:
100 points within $0.5 \le e < 1.0$,
100 points within $5.8{\rm\,yr} \le P < 6.3{\rm\,yr}$,
and initially 200 points for $T_0$ distributed over one orbital
period.
After the initial scan over $T_0$, the best estimate for $T_0$ was
improved by re-scanning a narrower range of $T_0$ centered on the
minimum that was found in the coarser scan.  This grid refinement was
repeated until the step size was less than one day.

\begin{figure}[t]
\centerline{\includegraphics[width=0.89\hsize]{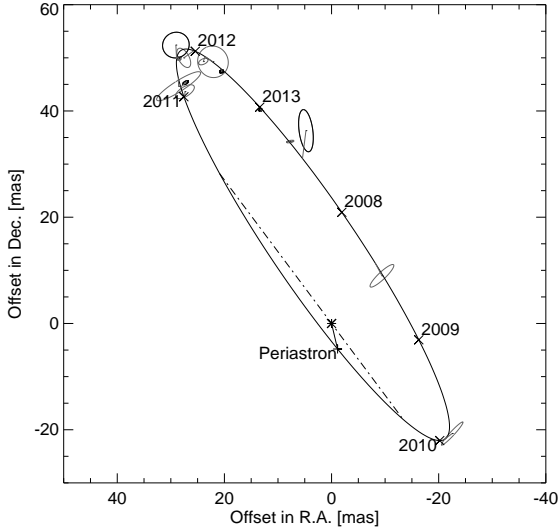}}
\caption{The orbit of component TWA\,5Ab around Aa.
  The observed positions are marked by their error ellipses and lines
  connecting the observed and calculated position at the time of the
  observations.
  The dash-dotted line indicates the line of nodes and the solid line
  the periastron.
  The crosses mark the expected positions at the beginning of the
  years 2008 to 2013.}
\label{OrbitPic}
\end{figure}

We improved the results of the grid-search with a Levenberg-Marquardt
$\chi^2$ minimization algorithm \citep{Press92} that fits for all 7
parameters simultaneously.  The simple approach would be to use
the orbital elements with the minimum $\chi^2$ found with the
grid-search.  However, initial test runs showed that the algorithm
does not converge on the global minimum.  For the same reason, we did
not use the previously published orbit solution as a starting point.
To make sure we find the global minimum value of $\chi^2$, we decided
to use all orbits resulting from the grid-search as starting points
that had $\chi^2 < \chi^2_{\rm min}+9$.
The number 9 was chosen arbitrarily to avoid starting from orbits that
are obviously bad.
The orbit with the global minimum $\chi^2$ found by the
Levenberg-Marquardt fit is shown in Fig.~\ref{OrbitPic}, and its
elements are listed in Table~\ref{fit2013_5_bestTab}.
To convert the semi-major axis from mas to AU, we used the
{\new new trigonometric distance of $50.1\pm1.8\rm\,pc$
  \citep{weinberger2012}}.

{\new We compute $\chi^2$ with the formula
$$
\chi^2 = \sum_i\left(
	\left({d_{{\rm obs},i}-d_{{\rm mod},i}\over\sigma_{d,i}}\right)^2 +
	\left({\rm PA}_{{\rm obs},i}-{\rm PA}_{{\rm mod},i}\over
		\sigma_{{\rm PA},i}\right)^2 \right),
$$
where $d$ and PA are separations and position angles\footnote{It
  is also possible to compute $\chi^2$ from the relative position in
  $x$ and $y$. This results in a different $\chi^2$ unless
  correlations between $x$ and $y$ are considered.  We expect
  $d$ and PA to be uncorrelated, since their errors are mostly caused
  by different effects (changes in pixel scale and rotation of the
  camera).}, respectively,
$\sigma_d$ and $\sigma_{\rm PA}$ are their respective errors, while
the indices, obs and mod, mark the observations and model
predictions.  The sum is computed over the observations.

The two rightmost columns in Table~\ref{ObsTab} list the differences
of $d$ and PA between the observations and the model.  Most of the
differences are less than $2\sigma$.  The biggest outlier is the
separation in January 2012.  We inspected the raw data and found that
the peak of the PSF was so bright that it reached into the non-linear
regime of the detector.  This might influence the position of the
center of light and therefore cause a bias in our measurement.
Therefore, we decided to omit the data point of January 2012 from our
orbit fit.  This changes the reduced $\chi^2$ from 3.5 to 2.1, the
semi-major axis from 3.1 to 3.2\,AU, and the system mass from 0.80 to
0.90\,$M_\odot$.  }

Errors of the orbital elements were determined by studying the
$\chi^2$ function around its minimum.
The uncertainty for each parameter corresponded to the point where
$\chi^2 = \chi^2_{\rm min} + 1$.
The reduced $\chi^2$ of our fit was 2.1, a bit more than one
would expect for a good fit.  This indicated that some of the errors
might have been underestimated.
To avoid underestimating the errors of the orbital elements as well,
we rescaled the errors of the observations so that the minimum
$\chi^2$ was 1.  Although some of the observations showed larger
deviations from the model than others, we had no reason to trust any
of the measurements less than the others.  Therefore, we multiplied
all observational errors by the same factor $\sqrt{2.1} = 1.45$.
The errors of the orbital elements in
Table~\ref{fit2013_5_bestTab} were based on $\chi^2$ computed
with these scaled measurement errors, while the deviations between
model and observations listed in Table~\ref{ObsTab} were computed with
the original measurement errors.

Estimating the error of the mass required a special procedure.
The mass itself was computed using Kepler's third law ($M=a^3/P^2$,
Kepler 1619).
The semi-major axis $a$ and the period $P$ are usually strongly
correlated.  To obtain a realistic estimate for the mass error,
we did {\em not\/} use standard error propagation.
Instead, we considered a set of orbital elements where the
semi-major axis was replaced by the mass.  This is possible because
Kepler's third law gives an unambiguous relation between the two
sets of elements.  With the mass being one of the orbital elements,
we treated it as one of the independent fit parameters and determined
its error in the same way as for the other parameters.


%
%
%
\begin{table}
\caption{Parameters of the best orbital solution.}
\label{fit2013_5_bestTab}
\renewcommand{\arraystretch}{1.3}
\begin{center}
\begin{tabular}{lr@{}l}
\noalign{\vskip1pt\hrule\vskip1pt}
Orbital Element				& \multicolumn{2}{c}{Value} \\
\noalign{\vskip1pt\hrule\vskip1pt}
Date of periastron $T_0$			& $2455328$ & $\,^{+   9}_{   -2}$\\
						& (2010 May 11)\span\\
Period $P$ (years)                              & $  6.025$ & $\,^{+0.010}_{-0.008}$\\
Semi-major axis $a$ (mas)                       & $   63.7$ & $\,^{+0.2}_{-0.2}$\\
Semi-major axis $a$ (AU)                        & $    3.2$ & $\,^{+0.1}_{-0.1}$\\
Eccentricity $e$                                & $  0.755$ & $\,^{+0.004}_{-0.003}$\\
Argument of periastron $\omega$ ($^\circ$)      & $  253.1$ & $\,^{+0.1}_{-0.2}$\\
P.A. of ascending node $\Omega$ ($^\circ$)      & $   36.5$ & $\,^{+0.3}_{-0.2}$\\
Inclination $i$ ($^\circ$)                      & $   97.5$ & $\,^{+0.1}_{-0.1}$\\
System mass $M_{Aa+Ab}$ ($\rm mas^3/year^2$)          & $   7130$ & $\,^{+ 129}_{ -57}$\\
Mass error from fit ($M_\odot$)			& $	$ & $^{+0.016}_{-0.007}$\\
Mass error from distance error ($M_\odot$)	& $	$ & $\pm 0.097$\\
System mass $M_{Aa+Ab}$ ($M_\odot$)                   & $   0.90$ & $\pm 0.1$\\ 
reduced $\chi^2$				& $    2.1$\\
\noalign{\vskip1pt\hrule\vskip1pt}
\end{tabular}
\end{center}
\end{table}


To check whether our error estimates for the orbital elements
  are reasonable, we employed the jackknife method.  We removed one
  epoch of our set of observations and repeated the orbit fit with the
  remaining data.  This was repeated for all observations with the
  exception of May 2002 and December 2003.  These two points are the
  sole measurements in the north- and southwestern sections of the
  orbits.  The standard deviations of the orbital elements resulting
  from this set of orbit fits are comparable in size to the error
  estimates obtained by $\chi^2$-analysis.
  The most important disparities are the errors of the semi-major
  axis, where the jackknife method yields an error of $\pm
  0.95\rm\,mas$, and the system mass with a fit-error of
  $\pm0.037\,M_\odot$.  However, if the error of the distance is
  included, the total error of the mass is $\pm0.1\,M_\odot$, which is
  the same as the error obtained by $\chi^2$-analysis.


{\new
\section{The orbit of TWA 5B and the mass ratio Aa/Ab}
\label{OrbitBsect}

The orbit of components TWA\,5Aa and Ab around each other allows us to
determine only the combined mass of TWA\,5Aa and Ab (which we will
abbreviate as Aa and Ab from now on).
To compute the individual masses, we need to know the mass ratio $q$,
which can be computed if the position of the center of mass (CM) of Aa
and Ab is known.  Unfortunately, we cannot observe the CM directly.
However, we know that TWA\,5B is in orbit around the CM of Aa and Ab
and that Aa and Ab are in orbit around their CM.
The CM is always on the line between Aa and Ab, and its distance from
Aa is the constant fraction\footnote{The parameter $f$ is often called
  fractional mass \citep{heintz78}, since it is the secondary star's
  fraction of the total mass in a binary.  It is useful in our case
  because it also describes the fractional offset of the CM from Aa,
  which is the separation Aa--CM divided by the separation Aa--Ab.},
$f=q/(1+q)$, of the separation of Aa and Ab.
With this information, the relative positions of TWA\,5Aa, Ab, and B
are sufficient to solve for the orbit of TWA\,5B and the mass ratio
Aa/Ab.


We follow the method that we used previously to derive individual
masses in the triple systems T~Tauri and LHS\,1070
\citep{Koehler2008,Koehler2012}.
The position of the CM of Aa and Ab is described in two ways:
First, it is on a Kepler-orbit around TWA\,5B, which is described
by seven orbital elements (For the mathematical description of the
Kepler-orbit, it is irrelevant whether we put the CM or TWA\,5B in the
center).
Second, the position of the CM can be computed from the observed
positions of Aa and Ab, and the mass ratio (which is treated as a free
parameter).
Standard error propagation is used to obtain an error estimate for
this position.  To compute $\chi^2$, we compare the position of the CM
from the orbit around TWA\,5B to the position derived from the
observations.
Our model has therefore eight free parameters: the seven elements,
which describe the orbit of the CM of Aa+Ab around B, and the
parameter $f=q/(1+q)$.
We chose to use the parameter $f$ instead of the mass ratio $q$, since
$f$ is confined to the range 0 to 1, while $q$ is a number between 0
and infinity.  Therefore, $f$ is better suited to a grid search.

We have only five two-dimensional measurements of the relative position
of TWA\,5B.  The PA changed by less than $2^\circ$ over the course of
the observations.  These data do not constrain the orbit very well.
Furthermore, the system mass of the best-fitting orbit model is
$6.4\,M_\odot$, which is much too high for TWA\,5.
To improve the situation, we also use our knowledge of the total
system mass.  Our orbit fit for TWA\,5Aa-b
yielded a combined mass of TWA\,5Aa and Ab of $0.9\pm0.1\,M_\odot$.
TWA\,5B is a brown dwarf \citep{webb1999}; its mass cannot
be higher than ${}\sim0.1\,M_\odot$.  To compute $\chi^2$ in our orbit
fit, we add a term of the form
$$
  \left( {M_{\rm mod} - 1\,M_{\odot} \over 0.3\,M_{\odot}} \right)^2,
$$
where $M_{\rm mod}$ is the total system mass of the model.  We adopted
a large error of $0.3\,M_\odot$ for our mass estimate, since we
want the orbit fit to be constrained mainly by the astrometric data.
We repeated the fit with mass errors of $1$, $3$, and $10\,M_\odot$ to
see how it affects the system mass determined by the fit results.
Only a mass error of $10\,M_\odot$ leads to a small change of the
system mass from $1.1$ to $1.3\,M_\odot$.  Changing the mass error had
no effect on the mass ratio.


\begin{figure}[t]
\centerline{\includegraphics[width=0.92\hsize]{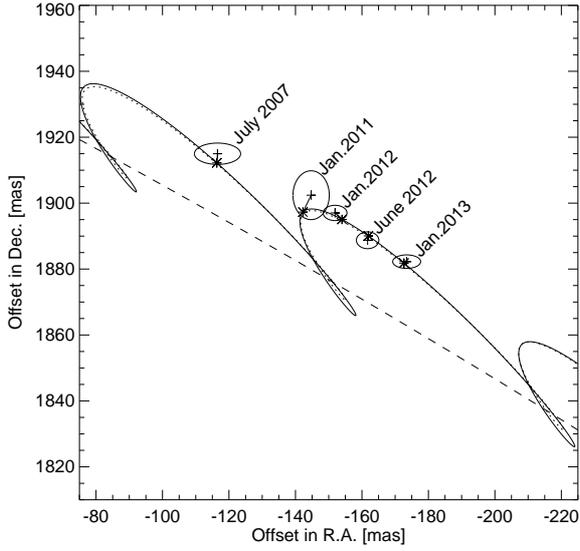}}
\caption{The motion of component TWA\,5B relative to Aa.
  The observed positions are marked by their error ellipses and lines
  connecting the observed and calculated position at the time of the
  observations.
  The dashed line is the Kepler-orbit around the center of mass of Aa
  and Ab.
  The solid line is our best fit for the motion of B relative to Aa.
  The dotted line is the best fit if the mass ratio Aa/Ab is fixed at
  1 or equal masses.
}
\label{OrbitABPic}
\end{figure}

%
%
%
\begin{table}
\caption{Parameters of the best orbital solution for A--B.}
\label{OrbABTab}
\renewcommand{\arraystretch}{1.3}
\begin{center}
\begin{tabular}{lr@{}l}
\noalign{\vskip1pt\hrule\vskip1pt}
Orbital Element				& \multicolumn{2}{c}{Value} \\
\noalign{\vskip1pt\hrule\vskip1pt}
Date of periastron $T_0$                       & $2513018$ & $\,^{+ 187}_{-187}$\\	
                                               & (2168 Apr 22)\span\\
Period $P$ (years)                             & $   1380$ & $\,^{+ 111}_{  -7}$\\	
Semi-major axis $a$ (mas)                      & $   2548$ & $\,^{+  23}_{  -7}$\\	
Semi-major axis $a$ (AU)                       & $    127$ & $\,^{+   4}_{  -4}$\\	
Eccentricity $e$                               & $   0.24$ & $\,^{+0.13}_{-0.10}$\\
Argument of periastron $\omega$ ($^\circ$)      & $    112$ & $\,^{+   5}_{  -4}$\\	
P.A. of ascending node $\Omega$ ($^\circ$)      & $   36.1$ & $\,^{+13.5}_{-0.6}$\\	
Inclination $i$ ($^\circ$)                      & $    138$ & $\,^{+  11}_{  -8}$\\	
System mass $M_{Aa+Ab+B}$ ($\rm mas^3/yr^2$)      & $   8685$ & $\,^{+  86}_{-170}$\\	
Mass error from fit ($M_\odot$)                 & $	$ & $^{+0.011}_{-0.021}$\\
Mass error from distance error ($M_\odot$)      & $	$ & $\pm 0.12$\\
System mass $M_{Aa+Ab+B}$ ($M_\odot$)             & $   1.1$ & $\pm 0.1$\\ 	
Mass ratio $M_{Ab}/M_{Aa}$                      & $    1.3$ & $\,^{+ 0.6}_{-0.4}$\\	
reduced $\chi^2$				&  & $    0.4$\\
\noalign{\vskip1pt\hrule\vskip1pt}
\end{tabular}
\end{center}
\end{table}


\begin{figure}[ht]
\centerline{\includegraphics[width=0.92\hsize]{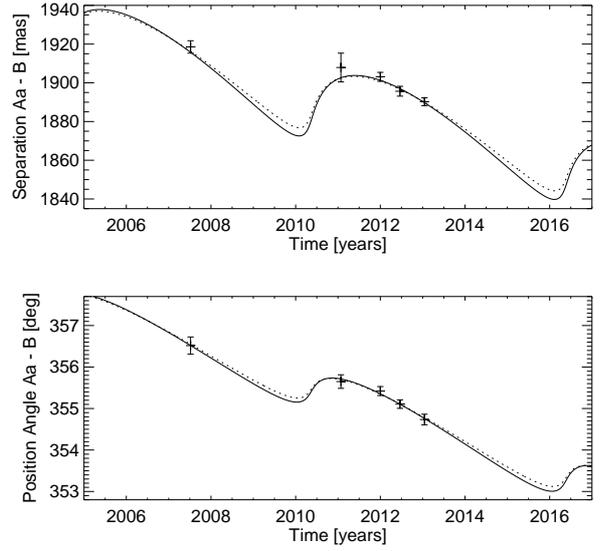}}
\caption{The motion of component TWA\,5B relative to Aa.
  Note that this is a combination of the orbit of B and the orbit of
  Aa, which are both around the CM of Aa/Ab.  The error bars are the
  measurements, the solid line is the best fit with the mass ratio
  Ab/Aa as a free parameter.  The dotted line shows the best-fitting
  orbit if the mass ratio is fixed at 1 or for equal masses of Aa and
  Ab.}
\label{SepPaAbPic}
\end{figure}


The fitting procedure is similar to that used for the orbit of Aa-Ab,
except that the grid-search is carried out in four dimensions:
eccentricity $e$, period $P$, time of periastron $T_0$, and the
fractional mass~$f$.
Singular value decomposition was used to fit the Thiele-Innes
constants, which give the remaining orbital elements.
It is worth noting that the orbital elements in this fit describe the
orbit of the A-B binary, only the fractional mass $f$ refers to the
pair Aa-Ab.

The four dimensional grid ranged from 0.3 to 0.7 in $f$, 0 to 0.25 in
$e$, and 100 to 3000 years in $P$.  The grid in $T_0$ started with 100
points distributed uniformly over one orbital period.  Similar to the
fit for the orbit of Aa-Ab, the grid in $T_0$ was refined until the grid
spacing was less than one day.

The parameters of the orbit with the globally minimum $\chi^2$ are
listed in Table~\ref{OrbABTab}.
  To estimate the errors, we employed the same method as in
  Sect.~\ref{InnerOrbitSect}.  We vary one parameter to find the
  point where $\chi^2 = \chi^2_{\rm min} + 1$.  Since $\chi^2_{\rm
    min}$ is smaller than 1, the measurement errors were used without
  scaling.
The motion of TWA\,5B relative to component Aa is shown in
Figs.~\ref{OrbitABPic} and \ref{SepPaAbPic}.

The reduced $\chi^2$ of 0.4 is surprisingly low, considering that the
fit is essentially based on the same data as the fit for the inner
orbit.  However, five astrometric positions correspond to ten
measurements, just two more than the eight parameters of the fit.
This means that our capability to detect inconsistencies in the data
is limited.

The mass ratio of 1.3 means that the component Ab is more massive than
Aa.  Since Aa is brighter, one would expect it to be more massive,
resulting in a mass ratio slightly below 1.
However, the uncertainties of our fit are large.
A mass ratio of 0.9 is within the {\newer 68.3\,\% confidence
  interval} of our fit.
The uncertainties are caused by the incomplete orbital coverage of our
observations (see Fig.~\ref{OrbitABPic}).  All our measurements of the
position of TWA\,5B have been collected when Ab was in the northeast
section of its orbit.
An observation in 2010, when Ab was in the southwest part of its
orbit, would have helped to measure the diameter of the astrometric
wobble of Aa around the CM.  Thanks to the period of Aa-Ab, this part
of the orbit will be reached again in 2015/2016.
} 


\begin{figure}[t]
\centerline{\includegraphics[angle=0,width=0.92\hsize]{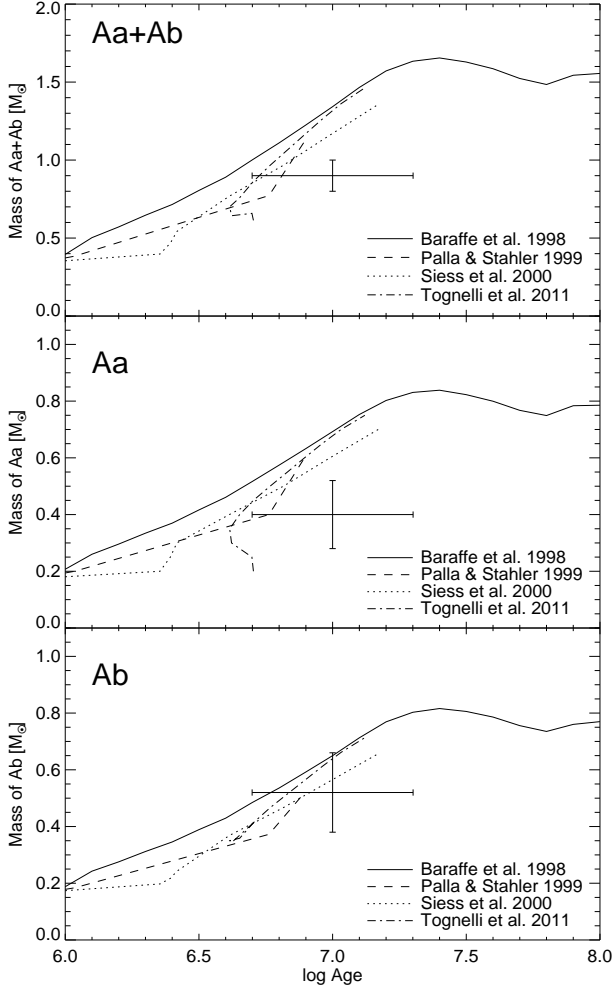}}
\caption{Plot of mass vs.\ age for the models with the same luminosity
  as TWA\,5Aa and Ab.
  The error bars indicate the measured age and system mass of TWA\,5A.
  The first panel shows the combined mass of Aa and Ab, while the
  second and third panel show the individual masses of Aa and Ab,
  respectively.
}
\label{MvsAgePic}
\end{figure}


\section{Comparison to theoretical models}

The aim of dynamical mass determinations is to test the predictions of
theoretical pre-main-sequence tracks.  The simplest way to do this
would be to pick a model with the measured mass, and compare the
predicted brightness at the age and distance of the object to the
measured brightness.  However, theoretical models follow the evolution
of individual stars, while our orbit yields the system mass, which is
the sum of the masses $M_{\rm Aa}+M_{\rm Ab}$.
{\new Individual masses computed with the mass ratio derived in
  Sect.~\ref{OrbitBsect} have larger uncertainties and are therefore
  less useful for comparing to theoretical predictions (see below).}

For now, we can go the other way: We take the resolved
{\newer bolometric luminosities \citep[who derived them from H-band
    magnitudes and bolometric corrections for PMS stars]{Konop2007}.
We compute absolute luminosities with the new trigonometric distance
  \citep{weinberger2012}.
Then, we find the corresponding theoretical models for
TWA\,5Aa and Ab, interpolating between tracks with different masses if
necessary.
Finally}, we compare the sum of the predicted masses to the dynamical
system mass derived here.

The first panel in Fig.~\ref{MvsAgePic} shows the system mass as
a function of age predicted by the models of \citet{BCAH98},
\citet{PallaStahler99}, \citet{siess2000}, and \citet{tognelli2011}.
Overplotted is the system mass of TWA\,5A resulting from our orbit
fit and the age of the TW Hydrae association.
The age is well constrained to be 5 -- 15\,Myr by a number of methods
\citep[and references therein]{weintraub2000}.

{\new
 {\newer At first glance,}
  all the models reproduce the data fairly well; most of them are
  within $1\,\sigma$.  The models of \citet{BCAH98} predict a younger
  age or higher mass but also reasonably agree with the data.

  The second and third panel in Fig.~\ref{MvsAgePic} show similar plots
  for the individual masses.  The mass of TWA\,5Aa is lower than
  predicted, while the mass of Ab is in good agreement with the
  models.  However, one should keep in mind that both masses are
  anti-correlated since the sum of both has to be
  $0.9\pm0.1\rm\,M_\odot$ (Note that the binary mass has been
  determined by the fit for the inner orbit and is in no way
  correlated with $q$).  Therefore, a mass of Aa at the upper end
  of the error bar corresponds to a mass $M_{\rm Ab}$ at the lower
  end, which would also be different from the model predictions.

 {\newer The models agree with the dynamical mass {\em only\/} if
   the age of TWA\,5A is at the young end of the age range for the
   TW~Hydrae association.  By placing TWA\,5A in an HR diagram,
   \citet{weinberger2012} found an age of 9 -- 11\,Myr.  With this
   age, the models predict a mass that is clearly higher than our
   dynamical mass.  The cause for this discrepancy is not clear, but
   we note that the HRD would give a younger age and lower masses if
   TWA\,5 was cooler than the temperature derived from its spectral
   type.  A temperature of $\sim 3500\,K$ would yield results that
   agree with the dynamical mass.}
}


\section{Conclusions and outlook}
\label{OutlookSect}

We present new relative positions of TWA\,5Aa and Ab and derive new
orbital elements from them.
{\new The system mass of $0.9\pm0.1\,M_\odot$ resulting from the new
  orbit is larger than the mass estimate previously obtained
  \citep{Konop2007} due to the larger distance based on the
  trigonometric parallax by \citet{weinberger2012}.
  With the new distance, the orbit of \citet{Konop2007} yields a mass
  of $1.0\pm0.2\,M_\odot$.}
The uncertainty of the mass determination resulting from the new orbit
fit is substantially smaller than before.
The uncertainty of the orbit fit is negligible compared to
the uncertainty of the distance.
{\new This situation will improve when the astrometric satellite
  {\it Gaia\/} delivers precise parallaxes.}

{\new We used the third component TWA\,5B as astrometric reference to
  measure the individual motion of TWA\,5Aa and Ab around their center
  of mass.  From this, we can derive the mass ratio of Aa and Ab and
  hence individual masses of the two components.
  Unfortunately, our relative measurements of TWA\,5Aa, Ab, and B
  cover only the northeast section of the inner orbit, resulting in
  large uncertainties for the mass ratio.
  It will reduce the errors significantly if the system is observed
  again in 2015/2016, when component Ab is in the southwest part of
  its orbit.

  A comparison of the dynamical mass(es) with those predicted by
  a number of theoretical models shows that all models considered are
  in reasonable agreement with the data.
  The errors of the mass and age determinations are still too large to
  distinguish between the models.
}

{\newer
  Dynamically, this system is an interesting case because it might be
  in Kozai resonance \citep{Kozai1962}.  This would explain the high
  eccentricity of the inner orbit.  An analysis of its dynamics will
  be published elsewhere (Beust et al., in prep.).

  As a secondary result, our data demonstrate that astrometric
  observations with NACO could be affected by a scatter in the
  instrument's orientation of 0.1 -- 0.2$^\circ$ with occasional
  larger jumps that can be explained by technical interventions.
  The cause for the random scatter could not be clarified with only
  one dataset at hand.  To guarantee high-precision astrometry, it
  would require astrometric standards being present within the
  observed field of view.  If no standards in the field are available,
  a calibration field should be observed shortly before or after the
  science target.
}

%
%
%

\bibliographystyle{bibtex/aa}
\bibliography{AA2012-20560}

\appendix

\section{Astrometric Calibration}
\label{AstromApp}

One of the more challenging tasks when doing astrometry with NACO is
the astrometric calibration, which is the determination of the precise
pixel scale and orientation of the detector at the time of the
observation.
During our astrometric monitoring program of TWA\,5 in 2011 to 2013,
images of the Hipparcos-binary HIP\,56620 (separation $3.390\arcsec$,
position angle $240.8^\circ$, \citealt{HIPPARCOS}) were taken.
To check whether the relative position of HIP\,56620 has changed in
the 20 years since the Hipparcos mission, we calibrated NACO with
a set of images of the Orion Trapezium cluster taken on 30\ January
2011.  Following the procedure described in Sect.~\ref{AstCalSect}, we
obtained a pixel scale of $(13.281\pm0.007)\rm\,mas/pixel$ and a PA of
$(0.80\pm0.02)^\circ$.
With this calibration, we find that the separation of
\object{HIP\,56620} on 27\ January 2011 was $(3.397\pm0.002)\arcsec$,
and the PA was $(240.27\pm0.12)^\circ$.
This position is significantly different from the position measured by
Hipparcos about 20 years earlier.

\begin{figure}[t]
\centerline{\includegraphics[angle=90,width=0.9\hsize]{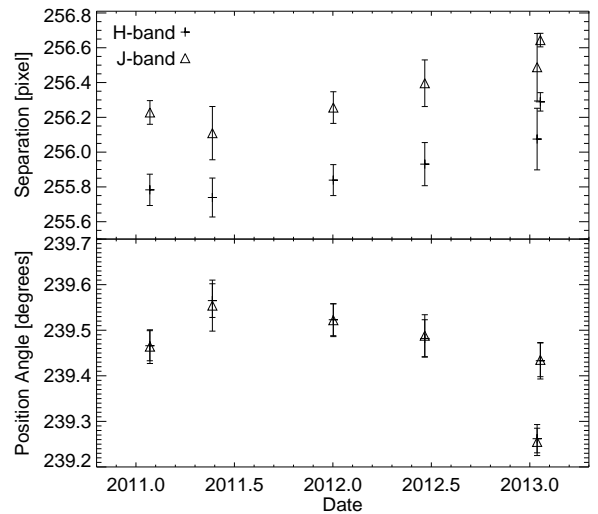}}
\caption{Relative position of the components of HIP\,56620 on the
  NACO detector.  \newer The position angle is in detector
  coordinates and measured counterclockwise from the $y$-axis.}
\label{CalibPic}
\end{figure}

Figure~\ref{CalibPic} shows separation and position angle of
HIP\,56620 from all our observations in detector coordinates
(separation in pixels, position angle measured counterclockwise from
the y-axis of the images).

First, we note that the separation measured in the J-band is always
larger by about 0.15\% than the separation measured in H.  It is
unlikely that the positions of the photocenter of the stars depend
on the wavelength.  Instead, this shift can be explained by chromatic
aberration of the optics in CONICA (R.\ Lenzen, priv.\ comm.).  This
means that the pixel scale of NACO is wavelength-dependent, albeit on
a scale that has to be considered only for high-precision astrometry.
To achieve the best precision, the pixel scale should
always be calibrated in the same filter as the observations of the
science target.

Second, we see a systematic change in both separation and position
angle between May 2011 and January 2013.  Orbital motion of the binary
might be an explanation, except that the point in January 2011 does
not follow the trend.  The orbital period is far too long to explain
this sudden change.

Third, HIP\,56620 was observed twice on 17 January 2013 due to the
scheduling of our observations by ESO.  These are two independent
observations with a new acquisition and optimization of the AO
system.  The separation measured in the two observations differs by
about 0.1\%, and the position angle by $0.17^\circ$.
There is no difference between the observations that could explain
this shift.  Therefore, we believe that this represents the intrinsic
precision of an astrometric observation of a binary with NACO.
Unless some sources that can be used for calibration are within the
field-of-view of the binary, the observations require moving the
telescope from the calibration field to the science target, which
causes changes of the calibration parameters that are similar to our
two observations of HIP\,56620 on 17 January 2013.


Since we do not know the true separation and position angle of
HIP\,56620 during our observations, we decided not to use it for the
astrometric calibration.  Instead, our calibration is based on two
data sets of the Trapezium cluster in Orion (Sect.~\ref{AstCalSect}).
Observations of clusters have the advantage that they contain more
than one star pair, which reduces the risk of undetected outliers.
As an estimate for the errors of the calibration parameters, we adopt
an error of $0.013\rm\,mas/pixel$ for the pixel scale and $0.1^\circ$
for the orientation.  This is larger than the scatter of the
calibration measurements but closer to the systematic uncertainties
of astrometric observations with NACO discussed in the previous
paragraph.

\end{document}